\documentclass[11pt]{article}

\usepackage{fullpage}

\usepackage{graphicx}
\usepackage{setspace}
\usepackage{xcolor}
\usepackage{bbm}
\usepackage{amsmath}
\usepackage{bbm}
\usepackage[hyphens]{url}

\usepackage{natbib}
\usepackage{booktabs}

\usepackage{hyperref} 
\hypersetup{
    colorlinks=true,    
    linkcolor=blue,     
    citecolor=blue,    
    urlcolor=blue
}

\usepackage{nicefrac}

\renewcommand{\paragraph}[2]{\vspace{2mm}\noindent \textbf{#2.}}

\title{Making AI Inevitable: Historical Perspective and the Problems of Predicting Long-Term Technological Change%
\footnote{The authors would like to thank Richard Ashcroft, Matthew Hamilton, Andrew Imbrie, Samuel Kimbriel, Daniel Losey for their helpful comments.
}}

\author{Mark Fisher%
\footnote{Assistant Professor, Georgetown University Department of Government,  
\texttt{mf1211@georgetown.edu}.}
\and John Severini%
\footnote{Ph.D. Student, Georgetown University Department of Government,
\texttt{jcs386@georgetown.edu}.}
}

\date{January 2025}

\begin{document}

\maketitle

This version is a pre-print of a now published article at \href{https://doi.org/10.1093/9780198945215.001.0001}{Oxford Intersections: AI in Society}.

It can be found at \href{https://doi.org/10.1093/9780198945215.003.0077}{doi.org/10.1093/9780198945215.003.0077}.

\begin{abstract}
This study demonstrates the extent to which prominent debates about the future of AI are best understood as subjective, philosophical disagreements over the history and future of technological change rather than as objective, material disagreements over the technologies themselves. It focuses on the deep disagreements over whether artificial general intelligence (AGI) will prove transformative for human society; a question that is analytically prior to that of whether this transformative effect will help or harm humanity. The study begins by distinguishing two fundamental camps in this debate. The first of these can be identified as "transformationalists," who argue that continued AI development will inevitably have a profound effect on society. Opposed to them are "skeptics," a more eclectic group united by their disbelief that AI can or will live up to such high expectations. Each camp admits further "strong" and "weak" variants depending on their tolerance for epistemic risk. These stylized contrasts help to identify a set of fundamental questions that shape the camps’ respective interpretations of the future of AI. Three questions in particular are focused on: the possibility of non-biological intelligence, the appropriate time frame of technological predictions, and the assumed trajectory of technological development. In highlighting these specific points of non-technical disagreement, this study demonstrates the wide range of different arguments used to justify either the transformationalist or skeptical position. At the same time, it highlights the strong argumentative burden of the transformationalist position, the way that belief in this position creates competitive pressures to achieve first-mover advantage, and the need to widen the concept of "expertise" in debates surrounding the future development of AI.
\end{abstract}

\section{Introduction}

In late May 2023, the Center for AI Safety put out a highly publicized “AI Extinction Statement” warning of the catastrophic risk involved in the continued development of artificial intelligence. The letter’s signatories brought together an overwhelming range of “AI experts,” including the eminently influential AI researcher Geoffrey Hinton, who quit Google that same month over fears about the future of AI.\footnote{At the time of the statement, Hinton was the most cited researcher within the fields of computer and cognitive science, as well as the second most cited in psychology (\citealp{rothman_why_2023}). While at the University of Toronto, Hinton was part of teams that made breakthrough discoveries leading to the AI boom of the 2010s. In the same capacity, he trained many other scientists who would be at the center of this same movement. Hinton left academia to work for Google in 2013. He was awarded the Turing Award in 2018 and the Nobel Prize in Physics in 2024.} The letter garnered an immediate response at the highest echelons of public life. Less than two weeks later, António Guterres, the UN Secretary General, proposed a sweeping set of relevant initiatives that sought to pave the way toward the global governance of AI. In doing so, he framed the effort as a direct response to the Center’s statement.

\begin{quote}
Alarm bells over the latest form of artificial intelligence—generative AI—are deafening, and they are loudest from the developers who designed it. These scientists and experts have called on the world to act, declaring AI an existential threat to humanity on par with the risk of nuclear war. We must take those warnings seriously (\citealp{guterres_secretary-generals_2023}).
\end{quote}

Guterres’ response captured the urgency of the “AI Extinction Statement.” Yet in presenting the statement’s message as the uniform assessment of experts, it also glossed over several deep divisions within the AI community. Far from sharing just one scientifically informed view on the future of AI, experts are deeply divided over two different, but overlapping questions. The most visible of these disagreements concerns how the achievement of artificial general intelligence (AGI)\footnote{Debates over the best way to characterize advanced AI are ongoing. In adopting the nomenclature of AGI, we do not have an overly precise definition in mind. Rather, we simply mean artificial intelligence that can match or surpass human-level intelligence across a range of different tasks (cf. \citealp{morris_position_2024, toner_oversight_2024}).} will affect human society, pitting those (such as the statement’s signatories) who fear it as an existential threat against those who believe it will be an unmitigated good. The second and more foundational question, though often the less visible, concerns whether we are in fact heading toward a world in which AGI proves transformative of human society.

In attempting to navigate these debates, policy leaders have often deferred to the AI expert community, a group that includes an overlapping set of computer scientists, tech leaders, and think tank researchers. There appear to be sound reasons for doing so. Practically speaking, any effective attempt to regulate the future development of AI will require the cooperation of not just nation-states and intergovernmental organizations but also of the corporations and scientific bodies at the forefront of such development (\citealp{pouget_future_2024, bremmer_ai_2023}). At an epistemic level, the AI community appears to be in an even more privileged position. Studies have repeatedly shown that the nontechnical public has a poor grasp of how AI technologies work (\citealp{sartori_minding_2023, arm_technology_ai_2020, long_what_2020, fast_long-term_2017}) and must therefore rely heavily on heuristics such as elite cues, metaphors, and other “folk theories” to navigate the increasingly AI-saturated world (\citealp{pohl_ai_2023, imbrie_contending_2021, ytre-arne_folk_2021, eslami_user_2019, french_whats_2017}). This would appear to place most political leaders and their constituents in a poor position to understand the technical issues involved in predicting the future trajectory of AI, and it has already resulted in widely divergent perceptions of AI’s transformative potential. As Helen Toner, a director at the Center for Security and Emerging Technology and former OpenAI board member, testified before the US Congress, “In public and policy conversations, talk of AGI is often treated as either a science-pipe dream or a marketing ploy” while “the scientists and engineers” of Big Tech pursue “it as an entirely serious goal” (\citealp{toner_oversight_2024}).

Yet it is easy to overstate the extent to which these more confident predictions can be effectively reduced to matters of technical expertise. Philosophers have noted that, given the still hypothetical nature of AGI, a detailed understanding of current AI technology will be insufficient to say anything with certainty about the profile of the AGI to come (\citealp{bostrom_future_2009}), and the community of AI experts has largely recognized that current disagreements over the potential risk of AGI stem from extra-technical sources. Hinton, for instance, has argued that the difference between techno-utopians and techno-pessimists is ultimately a matter of disposition and mood, not of technical disagreement—adding that his own concern over existential risk arises from the fact that he’s “mildly depressed” (\citealp{heaven_geoffrey_2023}). Others have been more cynical in their assessments. For Marc Andreessen (\citeyear{andreessen_why_2023}), the influential AI investor, techno-pessimists are simply “naive ideologues” or “self-interested opportunists,” the latter camp being made up of  “CEOs who stand to make more money if regulatory barriers are erected that form a cartel of government-blessed AI vendors protected from new startup and open source competition.” Alternatively, the prominent AI critics Timnit Gebru and Emile P. Torres (\citeyear{gebru_tescreal_2024}) have argued that techno-optimism and the particular brand of transhumanism it supports is grounded less in science than in a political ideology friendly to authoritarianism and second-wave eugenicism.

This study sets aside the question of whether AGI’s ultimate effect on society will be positive or negative. Instead, we focus on the extra-technical assumptions involved in predicting whether AI will prove transformative for society in the first place. We concentrate on this question partly due to its relative neglect. Despite being analytically prior to debates over AGI’s potential impact, it is routinely dropped from view in favor of the more attention-grabbing aspects of debating existential risk, as shown by Guterres’ statement. But we also believe that it makes for a more illuminating study of the role that non-technical assumptions play in predicting the future of technology more generally. Many scientific debates surrounding the development of AI concern highly technical questions related to resource availability, computational plateaus, and other problems concerning the ability of current technologies to scale up indefinitely, thereby justifying the nontechnical public’s deference to expert opinion. Policy debates, on the other hand, focus much more on the immediate term, centering around questions of risk management, adaptive governance, incident monitoring, uncertainty qualification, and model evaluation that are also firmly rooted in considerations of technical capacities. Yet, as this study will demonstrate, the divergence between these discussions, as well as within the broader public debate over the future of AI, is more fundamentally defined by a set of disagreements framed by contrasting perspectives on the historical process and underlying beliefs about technological change. This is a disagreement that takes place much more at the level of heuristics, metaphors, and “folk theories” than at that of technical knowledge (cf. \citealp{ogieblyn_god_2021}).

The primary aim of this study is to illuminate the differences between the various assumptions that underlie these perspectives, thereby helping both technical and nontechnical publics navigate the theoretical issues at hand. We begin by characterizing the debate’s two camps and offering a brief historical narrative of their development. The first of these we identify as the “transformationalists,” as they largely take for granted that AI development could soon achieve AGI and thereby alter the condition of society as we know it. Opposed to them are the “skeptics,” a more eclectic group united by their disbelief that AI can or will live up to its high expectations. Within both of these camps, we identify a further distinction between “strong” and “weak” variants depending on their approach to epistemic risk. In developing these stylized contrasts, we additionally identify a set of underlying questions that isolate the fundamental points of disagreement between the two camps and help to explain their respective interpretations of the future of AI’s development. We focus on three in particular: the possibility of non-biological intelligence, the appropriate time frame for technological predictions, and the assumed trajectory of technological development over the long term In highlighting these specific points of non-technical disagreement, this study demonstrates the wide range of different arguments that enter into justifying either the transformationalist or skeptical position. This demonstration serves to underline the formidable argumentative burden of the strong transformationalist position as well as the self-reinforcing implications of its adoption. It also suggests a need to widen the concept of “expertise” involved in debates over the future trajectory of AI.

\section{Transformationalists}

In their public roles as policy-relevant experts, transformationalists may seem like anything but a unified camp. In addition to representing competing tech companies, they often disagree openly over basic questions regarding when we will arrive at AGI, how we will do so, and whether this will come as a blessing or a curse. Yet, underlying these debates, transformationalists demonstrate a common set of intellectual commitments that serve to distinguish them from their skeptical counterparts. The most evident of these is the belief that AGI is on the horizon, and thus an eventuality that must be treated as a matter of great urgency. Propping up this assessment, however, is a much wider array of assumptions about the role of technology in society, the long-term patterns of technological progress, the constancy of historical change, and the subsequent directionality of the historical process.

In recognizing how all of these ideas come together, both historically and conceptually, the natural starting point is the Singularitarian movement. Despite eventually falling out of fashion, this movement has profoundly influenced the ways that many technical experts and futurists think about the prospects for AGI. It also remains a blueprint for those experts who are seriously trying to anticipate the future of technological progress more generally. At the core of Singularitarian thinking is the idea of the Singularity itself. Borrowed from mathematics and physics, this term was first invoked in the mid-twentieth century to hypothesize the point at which exponential technological progress approaches something resembling a vertical limit. Since then, however, idea has continued to evolve, resulting in no less than nine different overlapping theories of how to conceptualize the Singularity process itself (\citealp{more_overview_2013}). The most consequential of these theories are the “intelligence explosion” occasioned by superintelligent machines, the merger of human biology and machine intelligence, and/or the event-horizon created by such developments (cf. \citealp{yudkowsky_three_2007}).

In Singularitarian lore, the first person to theorize the Singularity was the mid-twentieth century polymath, John von Neumann, who is said to have noted in conversation that the “ever-accelerating progress of technology . . . gives the appearance of approaching some essential singularity in the history of the race beyond which human affairs, as we know them, could not continue” (\citealp{ulam_tribute_1958}). He was not the first to formalize this train of thought, however. As early as 1904, Henry Adams, the historian of technology and great-grandson of John Adams, theorized the implications of ever-accelerating technological progress in similar terms. Extrapolating from the doubling rate of coal extraction, Adams (\citeyear{adams_law_1904}) articulated a “Law of Acceleration” to account for what he saw as the long-term exponential growth of technological change. He eventually extended this law to “Thought” itself, which he believed to progress in a pattern as constant as any law of motion or electrodynamics. In drawing out the implications of these laws, Adams predicted that the rate of technological progress would resemble “a straight line to infinity” by or before the year 2025 (\citeyear[pp. 293, 305]{adams_rule_1909}).\footnote{Adams is not the only forerunner of Singularitarianism to be overlooked in most accounts of the movement. As Megan O’Gieblyn (\citeyear[pp. 43-77]{ogieblyn_god_2021}) has shown, there are also strong resonances—as well as a possible direct historical link—between the Singularitarians and the ideas of the French Jesuit Pierre Teilhard de Chardin.}

Singularitarianism emerged as a self-conscious movement in the early 1980s and 1990s, partly on the back of prevalent observations concerning the exponential trends in the development of computer hardware. In 1983, Vernor Vinge, a computer scientist and science fiction author, offered the first published statement asserting the inevitability of the coming technological singularity. In doing so, Vinge argued that we could be more certain of the Singularity’s eventuality than of the technical processes that would bring it about—a process which he claimed “cannot yet be predicted” but was also “not important” (\citealp{vinge_first_1983}). A decade later, however, Vinge (\citeyear{vinge_coming_1993}) elaborated on his theory in an influential paper for NASA, identifying three mechanisms of technological change that might be expected to lead toward the transcendence of humanity as we know it: a constant acceleration of technological progress, an eventual intelligence explosion, and an all too human desire for competitive advantage. This last mechanism was new to Vinge’s argument, which now included claims about the nature of human society. Given how much was to be gained by creating superintelligent machines, he argued, we could expect that people would stop at nothing to be the first to do so.

Vinge’s three mechanisms set the basic pattern for future transformationalist argumentation, both within and beyond those who explicitly identified as Singularitarians.\footnote{In line with this, there is a need to update David Chalmers’s (\citeyear{chalmers_singularity_2010}) claim that singularitarian thinking is defined by two theoretically distinct but practically compounding mechanisms, the “intelligence explosion” and the “speed explosion.” A third, the “competition explosion,” should also be included.}  At the same time, his work also set the stage for a divide within the transformationalist camp. Vinge argued that these mechanisms were sufficient to establish the inevitability of a radical, technology-driven transformation of humanity. We can call this teleological perspective “strong transformationalism.” Other transformationalist, however, have noted that this prospect is better characterized as a likely possibility. In qualifying Vinge’s argument, Nick Bostrom (\citeyear{bostrom_future_2009}), the former head of Oxford’s Future of Humanity Institute, noted that two conditions must be satisfied in order to achieve what he calls “posthumanity.” The first is that current trends in technological development continue into the indefinite future. The second is that superintelligence ultimately proves to be technologically feasible. Both of these conditions, he argues, are likely to hold true, but neither can be guaranteed a priori. We can call this more epistemically modest, probabilistic account “weak transformationalism.”

The most ambitious and influential attempt to shore up the strong transformationalist position has come from Ray Kurzweil (\citeyear{kurzweil_age_1990,kurzweil_singularity_2005, kurzweil_singularity_2024}), a decorated inventor, avowed Singularitarian, and “AI Visionary” at Google. Like Vinge before him, Kurzweil uses the exponential development of microchips described by “Moore’s Law” as a basic template for thinking about larger trends in the progress of technology. Unlike Vinge, however, he argues for the need to differentiate between the trajectory of any individual technology and the overall pattern of technological “evolution.” Individual technologies, he argues, will experience periods of exponential growth before leveling off into an S-curve, resulting in a plateau. At the same time, he argues that the overall shape of technological development is determined by the succession of technologies, each of which builds off those that came before it, not the trajectory of any individual technology. At the macro-level, this produces a “a cascade of S-curves,” each steeper than those before it, resulting in a global pattern of constant, exponential growth. Kurzweil styles this macro-pattern the “Law of Accelerating Returns” (\citealp[pp. 40-43]{kurzweil_singularity_2005}).

Kurzweil sees in this law much more than a likely account of modern technological progress. He extends it back to the very founding of the universe, arguing that it describes a steady “process of creating patterns of increasing order” that amounts to “the ultimate story of our world” (\citealp[p. 14]{kurzweil_singularity_2005}). Beginning with the first formation of basic chemical structures after the Big Bang, he argues, the law can be used to track an exponential growth in the complexity of intelligence throughout the universe.\footnote{Kurzweil (\citeyear[p. 388]{kurzweil_singularity_2005}) describes his basic metaphysical position as “patternism,” arguing that the underlying patterns that give enduring shape to ever-changing material phenomena represent the ultimate reality of the self and the universe. For a lucid discussion of Kurzweil’s patternism that places it in the longer tradition of speculation about the enduring properties of the self, see \citealt[pp. 43-78]{ogieblyn_god_2021}.}  On this curve, Kurzweil plots the formation of organic life, the emergence of brains capable of abstract thinking, the emergence of human technology, and, sometime in the mid-2040s, the Singularity, which he characterizes in terms of the fusion of human and artificial intelligence. Extrapolating even further, Kurzweil argues that intelligence will eventually “spread out from its origin on Earth” and “saturate” the universe by “reorganizing matter and energy to provide an optimal level of computation” (\citealp[p. 21]{kurzweil_singularity_2005}).

If Kurzweil’s Law of Accelerating Returns began as a reflection on the development of computer hardware, he eventually identifies it in terms of the intrinsic teleology of the universe. He describes the spread of order as the very “meaning” and “purpose” of the technological change we experience (\citealp[p. 7]{kurzweil_singularity_2005}). The Singularity is, he argues, our “destiny” (\citealp{kurzweil_singularity_2001}). In this way, Kurzweil is able to outflank the weak transformationalist’s skepticism and double down on the inevitability of transhuman superintelligence. It is nevertheless a step too far for many of his critics, who often accuse him of overstepping the boundaries of science and engaging in a form of religious argumentation (\citealp{beam_that_2005, gray_road_2011, bringsjord_belief_2012}; \citealp[p. 2]{bostrom_superintelligence_2014}; \citealp[pp. 43-77]{ogieblyn_god_2021}; \citealp[Section 4.2]{gebru_tescreal_2024}).

Kurzweil is routinely coy in the face of such claims.\footnote{For instance, when asked, “Does God exist?” in a documentary made about his life, Kurzweil responded, “Not yet” (\citealp{ptolemy_transcendent_2011}; \citealp[cf.][p. 389]{kurzweil_singularity_2005}).}  Other transformationalists, however, have approached the apparent religiosity of Singularitarianism with greater consternation. For some, like the anonymous creator of the infamous “Roko’s Basilisk” thought experiment, the prospect of a god-like superintelligence has presented itself as a matter of urgent speculation.\footnote{The idea, which has drawn parallels to both Pascal’s Wager (\citealp{singler_rokos_2018}) and Newcomb’s paradox (\citealp{auerbach_most_2014}), considers the circumstances in which an all-powerful and otherwise benign superintelligent agent might punish those who would not actively work towards the AI’s eventual realization.}  For others, the apparent mysticism of Kurzweil’s teleological interpretation of history has been seen as a reason to distance themselves from the movement’s characteristic language. Yet, in doing so, it’s not clear that either the strong or weak transformationalists have actually succeeded in letting go of the underlying assumptions of the Singularitarian perspective. 

Bostrom’s \textit{Superintelligence} (\citeyear{bostrom_superintelligence_2014}), a book that would hugely influence elite opinion around the existential risk of AI,\footnote{In addition to achieving New York Times bestseller status and earning Bostrom a profile in the New Yorker (\citealp{khatchadourian_doomsday_2015}), \textit{Superintelligence} has received explicit endorsements from the likes of Sam Altman, Bill Gates, Elon Musk, and Peter Thiel.}  offers an informative example of the way that teleological beliefs may seep into the arguments of weak transformationalists. The explicit purpose of \textit{Superintelligence} is to offer a rigorously argued perspective on how an all-powerful, superintelligent AI agent might emerge, behave, and ultimately threaten humanity. In establishing baseline expectations for the proximity of superintelligence, Bostrom overtly rejects the Singularitarian arguments of Vinge and Kurzweil. In their place, he turns to expert surveys. Based on these surveys, he concludes that “human-level machine intelligence” is likely to be “developed by mid-century,” if not “considerably sooner,” and that this might “soon thereafter result in superintelligence” (\citealp[p. 21]{bostrom_superintelligence_2014}).

Bostrom’s use of expert surveys to establish the prospects for AI’s developmental trajectory is well within the scope of respectable social scientific practice. However, given the pervasiveness of Singularitarian thinking within the AI community at that time—an influence possibly overrepresented in the surveys Bostrom uses\footnote{The most obvious concern when conducting a survey like this is the potential for unmeasurable selection effects. Participants inclined to respond to a survey asking for predictions on the timeline to AGI are almost certainly being selected on their underlying belief that AGI is a meaningful topic of concern in the first place, a problem noted by Müller \& Bostrom themselves (\citeyear[pp. 13-14]{muller_future_2016}). This is likely compounded by the low response rate in the more recent survey (31\%), as well as by the fact that respondents were primarily ($\sim$68\%) drawn from conferences run by the creators of the polls (\citeyear[pp. 2-4]{muller_future_2016}).}—it is likely that the predictions he uses to justify his concerns about superintelligence are fundamentally informed by Singulatarian thinking. Rather than escaping this influence, Bostrom’s use of expert surveys hides it from view and gives the predictions built upon it an air of scientific credibility. In this way, strong transformationalist arguments have continued to serve as the basis for the weak transformationalists’ calculation of the probability of achieving AGI.

The debt of more recent strong transformationalist thinking to Singularitarian argumentation is less subtle. Sam Altman, for instance, has articulated his unwavering belief in “Moore’s Law for everything,” shorthand for the idea that “technological progress follows an exponential curve” and will soon produce a future that is as “unimaginably great” as it is “unstoppable” (\citealp{altman_moores_2021}). Mustafa Suleyman has elaborated a similar perspective. A co-founder of DeepMind, he is (at the time of writing) simultaneously the CEO of Microsoft AI and a Senior Fellow at Harvard’s Belfer Center for Science and International Affairs, representing the closest thing that Big Tech has to a CEO-scholar. His \citeyear{suleyman_coming_2023} book, \textit{The Coming Wave}, like Bostrom’s \textit{Superintelligence}, tries to offer a realistic, rigorous analysis of the risks involved in further AI development. Also like Bostrom, Suleyman attempts to distance himself from overt affiliation with Singularitarianism. However, Suleyman does not run away from making strong claims about the inevitable transformation that AI will shortly bring. Rather, unlike Bostrom, he endorses a vision of the future where exponential growth is not open to doubt, rooting this belief in an autonomous logic of technological development.

The precise shape of this logic can nevertheless be hard to pin down with certainty. At times, Suleyman seems to suggest that exponential growth is simply an endogenous property of “technology.” Elsewhere, he adopts the analytical perspective of a contemporary social scientist to suggest that a series of exogenous factors might explain the inevitability of this progress.. At various points, he foregrounds considerations of technological demand, the cost of production, and the incentives of creators and CEOs. Yet, despite his frequent adoption of a more academic tone, Suleyman relies on the core set of strong transformationalist positions that have remained constant from Vinge forward: the exponential growth of computer hardware points to a larger pattern of exponential technological growth; this pattern can be expected to continue indefinitely; the inevitable result of this is runaway growth that we can barely understand and which will be almost impossible to control. Suleyman’s version of the argument thus avoids attributing an intrinsic teleology to the cosmos only by making “technology” itself into a Promethean monolith living out its own inevitable fate. A reified metaphor thus serves to replace an otherwise unpalatable metaphysics (cf. \citealp{ogieblyn_god_2021}).

\section{Skeptics}

Around the same time that Vinge was conceptualizing the technological singularity, philosophers of mind like John Searle and Colin McGinn had already begun to formalize both ontological arguments against the possibility of creating non-biological intelligence as well as epistemological arguments opposed to the idea that consciousness could be understood in any meaningful sense to begin with (\citealp{mcginn_character_1982, mcginn_can_1989, searle_minds_1980, searle_intentionality_1983, searle_minds_1986}). The concepts of “intelligence” and “consciousness” were so poorly theorized, they argued, that any attempt to construct them whole cloth was doomed to failure. Philosophical extrapolation from such a perspective was thus fundamentally flawed. In many ways, the specter of this criticism still lurks in the work done by those addressing the Singularity concept today. It is telling that, while philosopher Margaret Boden dedicates the final chapter of \textit{Artificial Intelligence: A Very Short Introduction} (\citeyear{boden_artificial_2018}) to the Singularity concept, she spends most of it criticizing the Singularitarians’ fundamental misunderstanding of the numerous obstacles that would have to surmounted before AI could advance to something akin to AGI. Moreover, she overtly describes herself as a skeptic not only of the Singularity vision but of the type of long-term thinking emblematic of transformationalists in the first place.

In aggregate, the more skeptical position is better thought of as an ongoing, collective reaction to the wide variety of arguments about the growth and change of technology in society than as an organized camp in any meaningful sense. As with the transformationalists, skeptics can also be conceived of having both “weak” and “strong” underlying epistemological orientations. Weak skeptics might best be thought of as those similarly interested in predicting long-term technological change, but who otherwise disagree about the specific rate of technological growth or the underlying models used for predictive purposes. Stronger skeptics, like Boden, believe that predictions about long-term technological growth cannot be meaningfully made or simply should not be prioritized when compared to short- or medium-term predictions due to the prevalence of non-trivial roadblocks in the way of current AI development. As a result, retrospectively sketching a single narrative telling of the skeptics runs the risk of streamlining a wide variety of perspectives largely arrived at idiosyncratically and with varying epistemological commitments. That said, even these skeptical viewpoints rely on a similar set of subjective beliefs about the nature of technology and the trajectory of human history. While such beliefs may be philosophically incompatible with those championed by the transformationalist camp, they are not necessarily based around either a singularly coherent or diametrically opposed ideological position.

The closest thing to a protagonist of the “anti-singularitarian” position has been Theodore Modis, whose early research on measuring complexity in large systems supplied much of the data that wound up in Kurzweil's \textit{The Singularity is Near} (\citealp{modis_forecasting_2002, modis_singularity_2006}). Modis' work, published primarily in the \textit{Technological Forecasting and Social Change} journal, often emphasizes the difference between exponential and sigmoid functions (s-curves) and the role this difference plays when conceptualizing the long-term history of technological and social development (\citealp{modis_strengths_2007}). Modis and others have presented arguments that largely hinge upon the claim that all phenomena subjected to competitive pressures have natural limits or face diminishing returns over time. As a result, they are better modeled by s-curves rather than exponential functions (\citealp{phillips_s-curves_2007}). The distinction between the two often serves as a prominent point of contention when skeptical and transformationalist proponents discuss their respective visions of the future. Indeed, the persistent rivalry between Kurzweil and Modis almost entirely centers around whether Kurzweil’s conceptualization of human progress fits onto an exponential model or merely represents the middle point of an s-curve. Modis has continued to argue that Kurzweil incorrectly applies an exponential growth model to many technological trends that are better thought of as following s-curves. Kurzweil, for his part, maintains that the two simply disagree about how long it will take for the s-curves to level off (\citealp{modis_why_2012}).\footnote{See Kurzweil’s appended response within the same section.}

That the debate between the two is entirely founded upon a disagreement regarding model specification rather than historical priors is emblematic of the field in general. There is ample evidence of the entrenchment of this model-oriented perspective towards technological change, even on the skeptical side. Indeed, Modis himself notes that he has probably carried out “the greatest number of logistic fits [s-curves]” of anyone in history (\citealp{modis_strengths_2007}). But this belief in the broad applicability of s-curves bears the same epistemological scrutiny as that of the exponential curve he and others criticize Kurzweil for employing. By characterizing s-curves as indicative of invariant, “natural” growth patterns, an extraordinarily wide variety of otherwise disaggregate processes are thus reduced to a single phenomenological class. Such an approach works well for predicting the carrying capacity of certain animal populations (an analogy he frequently calls upon), but it has a tougher time explaining a more diverse array of phenomena contingent on historical development.
Modis’ \citeyear{modis_end_2005} analysis of internet adoption ratesis illuminating in this respect. After fitting an s-curve onto the rate of adoption across the early 2000s, he anticipated that the total number of internet users would stabilize around 72\%. 

Today, the rate of internet use in the US hovers around 92\% and continues to increase. Modis readily admitted that his own prediction was meant to be taken as notional rather than explicitly scientific, and he recognized that future deviations of the data patterns could easily change the trajectory of the curve itself (\citealp{modis_end_2005}). But the uncertainty underlying the such a framing raises the question of the model’s generalizability. Assumptions of logistic growth patterns are appropriate when looking at an outcome with an inherent maximum like a population percentage, but applying these same assumptions to macro-historical processes requires that causally important events be both endogenously explained and exogenously provided. In this way, the model assumes what it tries to explain. These and other anomalies inevitably arise when fitting trends onto the grand historical record, and they hint at the problem of trying to fit a single model onto a wide variety of phenomena.

The “weak skeptic” position—of which Modis is perhaps exemplary—has continued to grow in the last few decades. One factor for this growth has been the increasing popularity of “big history,” a field that prioritizes situating human history within one or another longer patterns, cycles, or waves of progress, complexity, and change whose origins begin as early as the beginning of the universe. Similarly representative of this ever-present focus on identifying trends through large scale modeling are Andrey Korotayev and many of the contributors to the \textit{Journal of Big History}. Though less combative with the Singularity concept than are Modis and the journal of \textit{Technological Forecasting and Social Change}, the two groups’ epistemological commitments largely overlap in that their respective methodologies primarily revolve around attempts to apply quantitative analysis to measure change across human and cosmic history. Both also similarly recognize the epistemological tradeoff that comes with prioritizing broad, generalizable principles over the more context-specific analysis that characterizes most contemporary academic historical scholarship today.

The 2020 edited volume \textit{21st Century Singularity and Global Futures} exemplifies this approach as it relates to tackling the Singularity problem directly (\citealp{korotayev_21st_2020}). In a particularly telling chapter, systems theorist Graeme Donald Snooks overtly charges Kurzweil and other transformationalists with the crime of unscientific historicism, claiming that the models they use are insufficient for the conclusions they reach. To explain the patterns of human and cosmic development that the transformationalist address, Snooks argues a more dynamic model is (\citealp{snooks_is_2020}). The empirical trade-offs of long-term historical quantification are thus presumed to be acceptable given the scale of the analysis and the scope of big history’s potential implications. Characterized in this sense, weak skeptics like Snooks and Modis don’t necessarily disagree that the long-term historical quantification underlying the transformationalist position is epistemologically suspect; they simply believe that the derived conclusions about the future of technological development may be incorrect.

Contemporary macroeconomics serves as a particularly clear example of a more mainstream academic field interested in historical quantification. Many economists in the last few years have begun to directly address some of the questions raised about AI by the transformationalist camp, though across far shorter periods than the ones looked at by the big historians (\citealp{korinek_artificial_2017, berg_should_2018, agrawal_economics_2019, davidson_could_2021, trammell_economic_2023, korinek_scenarios_2024}). Here, AI is often conceptualized as an example of a general-purpose technology that has the ability to diffuse across an entire economy as opposed to just a handful of sectors. As a result, there has similarly been continued interest in estimating a timeframe for AI’s effects on various parts of society as well as any potential associated costs (\citealp{brynjolfsson_productivity_2021}). Positioned firmly in the language of traditional economics, however, most work here eschews terms like accelerating returns or intelligence explosions and instead focuses on the impact automation could have on inequality, wages, growth, and productivity.

The last of these measures, productivity - which roughly refers to the relative output for any given input, may be the closest available proxy for the idea of progress as conceptualized by those concerned with long-term history. Some economists conclude that AI may indeed have a transformative effect on productivity in the coming decades, especially if AGI is achieved (\citealp{davidson_could_2021, trammell_economic_2023, korinek_scenarios_2024}). Concerningly for the transformationalist position, however, much of the work directly connecting AI to productivity comes out of a small but burgeoning literature focused on identifying the causes of a slowdown in productivity growth across most industrialized nations over the past several decades (\citealp{autor_new_2024, goldin_why_2024}). There is some early evidence that this slowdown, which includes science and research, may partially stem from the continually increasing complexity and interconnectivity of the economy. Relatively lower innovation in specific sectors may produce innovation “bottlenecks.” These bottlenecks subsequently stymie growth across any sectors dependent upon complementary innovations from that bottlenecked sector, which in turn creates new and compounding bottlenecks as a result (\citealp{acemoglu_bottlenecks_2024}). Consider this in light of the 2020 paper, "Are Ideas Getting Harder to Find?," which has suggested that the marginal value of research has consistently gone down across a wide variety of sectors, leading to a precipitous increase in the cost of the very innovations needed to break through bottlenecks in the first place. The end of Moore’s Law is specifically highlighted as an indication of a general slowdown in technological development, but agricultural productivity and cancer research have also seen sharp decreases in their effective growth rates, which indicates that the trend could be more universal than previously thought. In aggregate, research productivity may in fact be halving approximately every 13 years, not doubling as would be expected by the transformationalists (\citealp{bloom_are_2020}).

If this implies that the low-hanging innovations may have already been plucked, then it could be the case that the rate of technological growth derived from AI development may soon start to slow down rather than continue to speed up. Accordingly, some have begun to systematically employ a distinction between "easy" and "hard" tasks for AI, arguing that productivity gains in field where outcomes are difficult to measure like management or design may be significantly more difficult to achieve than some of the more optimistic forecasts have thus far suggested (\citealp{mitchell_why_2021, noy_experimental_2023, peng_impact_2023, eloundou_gpts_2023}). If the distinction between easy and hard tasks is predicated on the availability of training data and if that training data merely lets AI replicate human behavior rather than productively augmenting it, then the snowball effect of continually self-improving AI anticipated by the transformationalists may be a mirage given its current methods of learning (\citealp{acemoglu_ai_2024}). Moreover, if AI ends up producing the bulk of information that it learns from, this cycle of replication can compound over time and lead to “model collapse,” or the poisoning of its own training data from the outset. Rather than learning from novel information drawn from human behavior, the AI will simply continue to repeat its own behaviors indefinitely (\citealp{shumailov_curse_2024}).\footnote{Acemoglu's substantive concerns about contemporary AI hew close to the type of criticism expressed by Searle and Boden: "When there is no good information on what the desired outcome is . . . most of the training data of AI models will come from how humans act in similar circumstances. As a result, learning from humans will not lead to better than human performance and is unlikely to generate new complementarities and reveal different insights than what humans are already doing." (\citeyear[p. 17]{acemoglu_ai_2024}).}

Epistemologically, the most notable takeaway from contemporary macroeconomics is the recognition of the deep uncertainty that comes with making longer-term estimates about the impact of technology, especially for a general-purpose technology like AI (\citealp{brynjolfsson_artificial_2019}). Although the philosophers of mind might stand on record as some of the oldest “strong” skeptics of the Singularity concept, many economists working on AI are beginning to voice similar concerns. Daron Acemoglu, for instance, deliberately limited his analysis of AI's potential future impact to no further than ten years out and included an expression of skepticism regarding any sort of longer-term forecasting about the potential impact AI might have on research productivity (\citealp{acemoglu_ai_2024}). Reminiscent of the older strong skeptics’ concerns, it comes as no surprise that his conclusions about AI’s impact fell far below that of even the “weak” skeptics, leading him to warn against believing the “AI hype” altogether (\citealp{acemoglu_dont_2024}).

\section{Key Questions in the AI Debate}

From these descriptions of the two camps, it is possible to identify three fundamental questions that get at the heart of the debate around AI’s impact on long-term technological change. Any position on the subject must provide an answer to all three, whether explicitly or assumed, yet in each case the answers on offer are not liable to scientific proof. These questions are as follows. First: is the technological development of a non-biological intelligence possible and can we know if we’ve achieved it? Second: what are appropriate time frames for making technological predictions? Third and finally: will the growth of innovation lead to a world of exponential technological development, or might technological development instead continue on linearly or enter into an eventual stagnation? By addressing  each question in turn, we are able to bring the various, non-technical assumptions that underlie disagreements about the future of  AI into sharper focus.

\subsection{Is technological development of a non-biological intelligence possible and can we know if we've achieved it?}

The first question revolves around whether genuine intelligence is ultimately feasible as an endpoint of technological development. If one believes that intelligence is an inherently biological product, as some of the skeptics have argued, then—by definition—artificial intelligence of any type is a misnomer. Strong skeptics like Searle and McGinn would offer the most direct ontological argument here: because intelligence is a phenomenon emergent only from biological evolution, it is likely that there are some inherent features of the biological process that make non-biological intelligence impossible. From this perspective, technological development will never result in or benefit from a uniquely non-biological intelligence explosion. Any increased ability of intelligence would only be possible through the enhancement of biological intelligence by non-biological augmentations.

As argued by Bostrom (\citeyear{bostrom_superintelligence_2014}), however, the potential uniqueness of biological intelligence does not preclude a superintelligence arrived at through biological evolution or modification. It does, however, entail a variety of limitations when compared to non-biological pathways. The vast difference in speed between transistors and human neurons speaks to this: the former are tens of millions of times faster than the latter. This implies that even if biological intelligence may be able to evolve into a sort of superintelligence over time, it will remain limited by the fundamental mechanics of biological processes. Artificial superintelligence might be an entirely different categorical concept.

If transistors can be used to replicate the role of biological neurons within a simulated, non-biological brain, then it raises the question of how we can meaningfully recognize intelligence outside of the biological context. Quantitative measures alone are unlikely to provide sufficient evidence of intelligence given that modern computers are capable of vastly superior capabilities than humans across a variety of straightforward but computationally-heavy domains. As indicated by contemporary AI’s highly varied substitutability across “easy” and “hard” tasks, there is clearly something more to intelligence than just the ability to optimize solutions for ideal, predesignated outcomes. 

It seems that the ability to meaningfully reason and comprehend is an altogether different process. Some of the currently understood limitations of AI, like the availability and use of training data, may be only incidentally problematic within a specific model paradigm and could pose less of a concern in the future. Other limitations, however, may be more fundamental to the nature of intelligence in the first place. If, as posed by the frame problem, intelligence requires the ability to distinguish contextually relevant from irrelevant information without the creation of an infinite number of axioms, then there may be more ambiguous thresholds that non-biological intelligence has yet to surmount.\footnote{For more on the frame problem, see \citealt{dennett_cognitive_1984, shanahan_frame_2004}.}

Intelligence in this sense, then, is not so much an objective, measurable phenomenon as much as it is an abstract concept beholden to the subjective beliefs of the observer. But if artificial intelligence is unable to surmount some of the fundamental obstacles laid before it, then both the nature of the intelligence explosion and the timeline of any resultant accelerating returns remain up for debate.

\subsection{What are appropriate time frames for making technological predictions?}

The second question deals with the overall willingness to scrutinize a significantly longer scope of history that looks back to at least the beginning of humanity (and often all the way to the Big Bang) in order to derive trends and predictions about the future of technological development. Generally speaking, the type of work conducted at this level of abstraction might be most accurately described as pre-paradigmatic in a Kuhnian sense, as it currently seems to lack a cohesive framework for empirical analysis. This often seems to lead to intractable arguments, as illustrated by the Kurzweil-Modis dispute or the methodological criticism posed by Snooks.\footnote{Though it is clear that this problem has been recognized for a while now. At least some have outwardly called for greater attention to be paid to methodology and concepts found in philosophy of science, however. See, for instance, \citealt{ho_typology_2015, phillips_key_2016}.}

These problems can partially be blamed on the longstanding indifference that academia–especially the humanities and social sciences–has had towards the enduring discussions surrounding long-term technological change, artificial intelligence, and big history more generally (\citealp{christian_case_1991}). As a result, the groundwork was laid for the development of an idiosyncratic methodological toolkit used to interpret history across such a long time frame that remains at least partially incommensurate with the toolkit employed by social scientists looking at shorter scopes. Notable in this sense are the plethora of unique terminology often employed but rarely overlapping with concepts in philosophy of science or history. Notions like "accelerating change," or "phases of history" are commonplace in works looking at big history but are infrequently employed in the work of more mainstream academics focusing on smaller historical scope (cf. \citealp{lepoire_exploring_2021}). Attempts to bridge this gap and employ both perspectives can thus invite internal tension, as seen in Suleyman’s \textit{The Coming Wave} (\citeyear{suleyman_coming_2023}).

As noted by Ben Goertzel in 2007, there exists a marked academic bias against interacting with especially long-term historical perspectives like those proposed by Ray Kurzweil. Goertzel observed that "the vast majority of academic and industry AI researchers remain deeply skeptical of the sort of optimistic rhetoric and perspective that Kurzweil's book typifies" (\citealp[p. 1162]{goertzel_human-level_2007}). This skepticism towards even addressing the subject in the first place has resulted in a deep tension that has undeniably shaped the contours of the debate about AI's long-term potential effects on humanity. The hardest skeptics like Boden or Acemoglu often do come from more mainstream academia and are frequently unwilling to engage in conversations around AGI or the Singularity because such subjects require for them too great an epistemological leap to warrant consistent inquiry. Alternatively, thinkers like Bostrom accept the methodology and empirical trade-offs requires to seriously participate in the ongoing debates about AI, superintelligence, and the Singularity concept.

\subsection{Is a world of exponential growth more likely than a world of steady-state growth or stagnation?}

Finally, there is the question about the trajectory of technological development more generally. Referring back to contemporary macroeconomics here provides a useful, if limited lens in formalizing the concept. Imbued across the broadest conceptualization of human history, this synthetic notion of long-term technological development could be imagined as a sort of aggregate growth of productivity. This in turn can be more explicitly thought of as the continued ability to overcome innovation bottlenecks through the successful creation of new ideas and the subsequent implementation and diffusion of corresponding innovations (\citealp{acemoglu_bottlenecks_2024, bloom_are_2020}).

A few basic underlying assumptions should be stated here: (1) The rate of bottlenecks and innovations would likely be thought of as inherently proportional to some level of aggregate technological development, all else being equal. (2) New bottlenecks would likely not be thought of as either adding to or subtracting from aggregate technological development. (3) New innovations would likely add to aggregate technological development, presumably in proportion to the total aggregate technological development already present. (4) New bottlenecks and the corresponding innovations that could break through them both would likely be thought of as requiring increasing amounts of aggregate technological development to recognize and create, respectively.

These assumptions form the basis for thinking about the trajectory of technological development and human progress across a long-term perspective. Accordingly, a simple tripartite model of possible futures emerges when considering the relationship between the growth of bottlenecks and innovations. The three worlds can be thought of as ideal points representative of the different expectations for the future of technological progress. Stated as expectations of the relative growth of (B)ottlenecks and (I)nnovations, they are:

\begin{enumerate}
    \item \underline{Stagnation} (B $>$ I): Where bottlenecks become more numerous and more difficult to deal with as the complexity of the world increases. This has a negative compounding effect on the creation of new innovations, which begin to require continually greater amounts of time and effort to successfully create and diffuse.
    \item \underline{Steady-state} (B $=$ I): Where bottlenecks and innovations remain roughly in equilibrium. Bottlenecks may become more numerous and more difficult to deal with, but new ideas keep pace, allowing for a relatively proportional rate of the creation and diffusion of new innovations.
    \item \underline{Singularity} (B $<$ I): Where bottlenecks are still likely to become more numerous and more difficult to deal with, but where new ideas and innovations enable relatively faster growth. This has a positive compounding effect on the creation of new innovations, which become easier to create and diffuse as a result of previous innovations continually lowering the costs of subsequent technological development.
\end{enumerate}

The intelligence explosion concept and Kurzweil’s theory of accelerating returns both correspond with the Singularity world, as the creation of new innovations leads to an inevitable snowball effect of self-improving technological development. Skeptics like Modis would contest the inevitability of such a world given the different, non-teleological trajectories that could occur in the future as a result of unforeseen contingencies like war or economic crisis. Most pessimistically, it may very well be the case that the exponential rate of technological development over the last centuries is anomalous and will begin to level off as the 21st century continues on with many bottlenecks but few innovations, leading to a Stagnation world. Finally, technological development might occur in cycles oscillating between both exponential and logarithmic moments or remain roughly linear over a long period of time. As typified by the skepticism of the contemporary macroeconomists like Acemoglu, the trajectory of human development will be difficult to predict, but will likely involve moments of both slowing down and speeding up as bottlenecks and innovations trade off with one another, leading to a Steady-state world. These ideal type worlds need not be permanent states of affairs, but the belief in the likely trajectory of technological development within at least the near future is a crucial underlying element of the AI debate today.

\section{Conclusion}

The three preceding questions help draw out the foremost beliefs underlying discussions regarding AI’s potential impact on society. They highlight the fundamental differences between the transformationalist and skeptic positions while demonstrating that the debate between the two cannot simply be reduced to questions of empirics. At the same time, however, they also stress the high argumentative burden faced by the transformationalist position – a burden perhaps underappreciated given its prevalence among both technical experts and lay populations alike. To justify the belief that AGI is on the horizon, one must argue that non-biological intelligence is ultimately feasible and identifiable, that we have good reasons to believe that the historical process remains relatively unchanged over the long term, and that technology will continue to progress at an exponential rate. To assert the strongest transformationalist perspective, one must furthermore argue that we can be absolutely sure of these positions. Such arguments ultimately rely on an epistemological perspective that more readily incorporates a sort of faith in technology than it does anything resembling science.

Difficult as such arguments may be, they are not impossible, especially in their weaker form. Our aim is not so much to delegitimize the transformationalist perspective as it is to demonstrate the complexity of the questions surrounding the prediction of long-term technological development. Nevertheless, the high burden of the transformationalist position bears emphasizing on account of the self-reinforcing effects of adopting this belief. As Vinge first noticed, the competitive pressures for first-mover advantage among those who do believe that AGI is inevitable are immense, encouraging both states and corporations alike to stop at nothing to bring it about. If the achievement of AGI does in fact become an inevitability, it will be in large part because enough powerful competitors within the AI space have come to believe it is so. An adequate appreciation of the difficulty of approaching such a conclusion with confidence, on the other hand, helps to create space within which governments, corporations, and other stakeholders can assert some degree of agency over the future development of the technology.

The difficulty of doing justice to complexity of predicting future technological advancements speaks to the persistent problems of theoretical speculation in the AI space. Indeed, it emphasizes the importance of broadening the range of expertise called upon to inform the political and social responses to the ongoing development of AI. Such a range should extend well beyond the requisite scientific and regulatory expertise more commonly prioritized today. The debates about AI and its effects on society need to incorporate far more philosophers of history, science, and technology, epistemologists, and political theorists. As we show, these debates do not revolve around technical questions with scientific answers, but instead around questions that may never be fully answerable in the first place.

Yet this is not simply to say that the balance of power ought to be tilted back towards academic humanists and social scientists without any further qualification. If transformationalist arguments might be criticized for lack of due rigor or idiosyncratic language, the modern academy holds at least partial responsibility. Questions about the longue durée of history and technological development are often treated with dismissiveness within the disciplines that are best positioned to address them, especially when these questions are perceived to come from members of the tech industry outside the ivory tower. As a result, the debates surrounding these questions have been forced to occur without the more conservative, disciplining effects of academic discussion and publication. Yet they are debates that play an important role in determining the social and political response to what is now a political and social issue of the most pressing variety, regardless of whether AGI is ever ultimately achieved. This study hopes to show one way in which this wider range of academics stand to contribute to informing the collective response to these issues if they are willing to take them up directly. We hope scholars become more willing to rise to the challenge.

\bibliographystyle{abbrvnat}
\bibliography{MakingAIInevitable.bib}

\begin{thebibliography}{83}
\providecommand{\natexlab}[1]{#1}
\providecommand{\url}[1]{\texttt{#1}}
\expandafter\ifx\csname urlstyle\endcsname\relax
  \providecommand{\doi}[1]{doi: #1}\else
  \providecommand{\doi}{doi: \begingroup \urlstyle{rm}\Url}\fi

\bibitem[Acemoglu(2024{\natexlab{a}})]{acemoglu_ai_2024}
D.~Acemoglu.
\newblock The {AI} {Safety} {Debate} {Is} {All} {Wrong} {\textbar} by {Daron} {Acemoglu}.
\newblock \emph{Project Syndicate}, Aug. 2024{\natexlab{a}}.
\newblock URL \url{https://www.project-syndicate.org/commentary/ai-safety-human-misuse-more-immediate-risk-than-superintelligence-by-daron-acemoglu-2024-08}.
\newblock Section: Innovation \& Technology.

\bibitem[Acemoglu(2024{\natexlab{b}})]{acemoglu_dont_2024}
D.~Acemoglu.
\newblock Don’t {Believe} the {AI} {Hype}.
\newblock \emph{Project Syndicate}, May 2024{\natexlab{b}}.
\newblock URL \url{https://www.project-syndicate.org/commentary/ai-productivity-boom-forecasts-countered-by-theory-and-data-by-daron-acemoglu-2024-05}.
\newblock Section: Innovation \& Technology.

\bibitem[Acemoglu et~al.(2024)Acemoglu, Autor, and Patterson]{acemoglu_bottlenecks_2024}
D.~Acemoglu, D.~Autor, and C.~Patterson.
\newblock Bottlenecks: {Sectoral} {Imbalances} and the {US} {Productivity} {Slowdown}.
\newblock \emph{NBER Macroeconomics Annual}, 38:\penalty0 153--207, Apr. 2024.
\newblock ISSN 0889-3365.
\newblock \doi{10.1086/729196}.
\newblock URL \url{https://www.journals.uchicago.edu/doi/abs/10.1086/729196}.
\newblock Publisher: The University of Chicago Press.

\bibitem[Adams(1904)]{adams_law_1904}
H.~Adams.
\newblock A {Law} of {Acceleration}.
\newblock In \emph{The {Education} of {Henry} {Adams}}. Oxford University Press, 1904.
\newblock ISBN 978-0-19-198990-2.
\newblock URL \url{https://www.bartleby.com/lit-hub/the-education-of-henry-adams/a-law-of-acceleration-1904/}.
\newblock Section: The Education of Henry Adams.

\bibitem[Adams(1909)]{adams_rule_1909}
H.~Adams.
\newblock The {Rule} of {Phase} {Applied} to {History}.
\newblock In \emph{The {Degradation} of the {Democratic} {Dogma}}. Macmillan, 1909.
\newblock URL \url{https://ericsteinhart.com/progress/adams-phase.pdf}.

\bibitem[Agrawal et~al.(2019)Agrawal, Gans, and Goldfarb]{agrawal_economics_2019}
A.~Agrawal, J.~Gans, and A.~Goldfarb, editors.
\newblock \emph{The {Economics} of {Artificial} {Intelligence}: {An} {Agenda}}.
\newblock University of Chicago Press, Chicago, May 2019.
\newblock ISBN 978-0-226-61333-8.

\bibitem[Altman(2021)]{altman_moores_2021}
S.~Altman.
\newblock Moore's {Law} for {Everything}, Mar. 2021.
\newblock URL \url{https://moores.samaltman.com/}.

\bibitem[Andreessen(2023)]{andreessen_why_2023}
M.~Andreessen.
\newblock Why {AI} {Will} {Save} the {World}, June 2023.
\newblock URL \url{https://a16z.com/ai-will-save-the-world/}.

\bibitem[Arm(2020)]{arm_technology_ai_2020}
Arm.
\newblock \emph{{AI} today, {AI} tomorrow: the arm 2020 global {AI} survey}.
\newblock Northstart, 2020.

\bibitem[Auerbach(2014)]{auerbach_most_2014}
D.~Auerbach.
\newblock The {Most} {Terrifying} {Thought} {Experiment} of {All} {Time}.
\newblock \emph{Slate}, July 2014.
\newblock URL \url{https://slate.com/technology/2014/07/rokos-basilisk-the-most-terrifying-thought-experiment-of-all-time.html}.

\bibitem[Autor et~al.(2024)Autor, Chin, Salomons, and Seegmiller]{autor_new_2024}
D.~Autor, C.~Chin, A.~Salomons, and B.~Seegmiller.
\newblock New {Frontiers}: {The} {Origins} and {Content} of {New} {Work}, 1940–2018*.
\newblock \emph{The Quarterly Journal of Economics}, 139\penalty0 (3):\penalty0 1399--1465, Aug. 2024.
\newblock ISSN 0033-5533.
\newblock \doi{10.1093/qje/qjae008}.
\newblock URL \url{https://doi.org/10.1093/qje/qjae008}.

\bibitem[Beam(2005)]{beam_that_2005}
A.~Beam.
\newblock That {Singularity} {Sensation}.
\newblock \emph{The Boston Globe}, Feb. 2005.

\bibitem[Berg et~al.(2018)Berg, Buffie, and Zanna]{berg_should_2018}
A.~Berg, E.~F. Buffie, and L.-F. Zanna.
\newblock Should we fear the robot revolution? ({The} correct answer is yes).
\newblock \emph{Journal of Monetary Economics}, 97:\penalty0 117--148, Aug. 2018.
\newblock ISSN 0304-3932.
\newblock \doi{10.1016/j.jmoneco.2018.05.014}.
\newblock URL \url{https://www.sciencedirect.com/science/article/pii/S0304393218302204}.

\bibitem[Bloom et~al.(2020)Bloom, Jones, Van~Reenen, and Webb]{bloom_are_2020}
N.~Bloom, C.~I. Jones, J.~Van~Reenen, and M.~Webb.
\newblock Are {Ideas} {Getting} {Harder} to {Find}?
\newblock \emph{American Economic Review}, 110\penalty0 (4):\penalty0 1104--1144, Apr. 2020.
\newblock ISSN 0002-8282.
\newblock \doi{10.1257/aer.20180338}.
\newblock URL \url{https://pubs.aeaweb.org/doi/10.1257/aer.20180338}.

\bibitem[Boden(2018)]{boden_artificial_2018}
M.~A. Boden.
\newblock \emph{Artificial {Intelligence}: {A} {Very} {Short} {Introduction}}.
\newblock Oxford University Press, Oxford, UK, Dec. 2018.
\newblock ISBN 978-0-19-960291-9.

\bibitem[Bostrom(2009)]{bostrom_future_2009}
N.~Bostrom.
\newblock The {Future} of {Humanity}.
\newblock \emph{Geopolitics, History, and International Relations}, 1\penalty0 (2):\penalty0 41--78, 2009.

\bibitem[Bostrom(2014)]{bostrom_superintelligence_2014}
N.~Bostrom.
\newblock \emph{Superintelligence: {Paths}, {Dangers}, {Strategies}}.
\newblock Oxford University Press, Oxford, UK, 2014.
\newblock ISBN 978-0-19-967811-2.

\bibitem[Bremmer and Suleyman(2023)]{bremmer_ai_2023}
I.~Bremmer and M.~Suleyman.
\newblock The {AI} {Power} {Paradox}.
\newblock \emph{Foreign Affairs}, 102\penalty0 (5):\penalty0 26--43, Oct. 2023.

\bibitem[Bringsjord et~al.(2012)Bringsjord, Bringsjord, and Bello]{bringsjord_belief_2012}
S.~Bringsjord, A.~Bringsjord, and P.~Bello.
\newblock Belief in {The} {Singularity} is {Fideistic}.
\newblock In A.~H. Eden, J.~H. Moor, J.~H. Søraker, and E.~Steinhart, editors, \emph{Singularity {Hypotheses}: {A} {Scientific} and {Philosophical} {Assessment}}, pages 395--412. Springer, Berlin, Heidelberg, 2012.
\newblock ISBN 978-3-642-32560-1.
\newblock \doi{10.1007/978-3-642-32560-1_19}.
\newblock URL \url{https://doi.org/10.1007/978-3-642-32560-1_19}.

\bibitem[Brynjolfsson et~al.(2019)Brynjolfsson, Rock, and Syverson]{brynjolfsson_artificial_2019}
E.~Brynjolfsson, D.~Rock, and C.~Syverson.
\newblock Artificial {Intelligence} and the {Modern} {Productivity} {Paradox}: {A} {Clash} of {Expectations} and {Statistics}.
\newblock In \emph{1. {Artificial} {Intelligence} and the {Modern} {Productivity} {Paradox}: {A} {Clash} of {Expectations} and {Statistics}}, pages 23--60. University of Chicago Press, June 2019.
\newblock ISBN 978-0-226-61347-5.
\newblock \doi{10.7208/9780226613475-003}.
\newblock URL \url{https://www.degruyter.com/document/doi/10.7208/9780226613475-003/pdf?licenseType=restricted}.

\bibitem[Brynjolfsson et~al.(2021)Brynjolfsson, Rock, and Syverson]{brynjolfsson_productivity_2021}
E.~Brynjolfsson, D.~Rock, and C.~Syverson.
\newblock The {Productivity} {J}-{Curve}: {How} {Intangibles} {Complement} {General} {Purpose} {Technologies}.
\newblock \emph{American Economic Journal: Macroeconomics}, 13\penalty0 (1):\penalty0 333--372, Jan. 2021.
\newblock ISSN 1945-7707, 1945-7715.
\newblock \doi{10.1257/mac.20180386}.
\newblock URL \url{https://pubs.aeaweb.org/doi/10.1257/mac.20180386}.

\bibitem[Chalmers(2010)]{chalmers_singularity_2010}
D.~J. Chalmers.
\newblock The {Singularity}: {A} {Philosophical} {Analysis}.
\newblock \emph{Journal of Consciousness Studies}, 17\penalty0 (9-10):\penalty0 7--65, 2010.
\newblock Publisher: Imprint Academic.

\bibitem[Christian(1991)]{christian_case_1991}
D.~Christian.
\newblock The {Case} for "{Big} {History}".
\newblock \emph{Journal of World History}, 2\penalty0 (2):\penalty0 223--238, 1991.
\newblock ISSN 1045-6007.
\newblock URL \url{https://www.jstor.org/stable/20078501}.
\newblock Publisher: University of Hawai'i Press.

\bibitem[Davidson(2021)]{davidson_could_2021}
T.~Davidson.
\newblock Could {Advanced} {AI} {Drive} {Explosive} {Economic} {Growth}?
\newblock Technical report, Open Philanthropy, June 2021.
\newblock URL \url{https://www.openphilanthropy.org/research/could-advanced-ai-drive-explosive-economic-growth/}.

\bibitem[Dennett(1984)]{dennett_cognitive_1984}
D.~C. Dennett.
\newblock Cognitive {Wheels}: {The} {Frame} {Problem} of {AI}.
\newblock In C.~Hookway, editor, \emph{Minds, {Machines} and {Evolution}}, pages 1--16. Bean, 1984.

\bibitem[Eloundou et~al.(2023)Eloundou, Manning, Mishkin, and Rock]{eloundou_gpts_2023}
T.~Eloundou, S.~Manning, P.~Mishkin, and D.~Rock.
\newblock {GPTs} are {GPTs}: {An} {Early} {Look} at the {Labor} {Market} {Impact} {Potential} of {Large} {Language} {Models}, Aug. 2023.
\newblock URL \url{http://arxiv.org/abs/2303.10130}.
\newblock arXiv:2303.10130 [cs, econ, q-fin].

\bibitem[Eslami et~al.(2019)Eslami, Elazari Bar~On, Vaccaro, Gilbert, Lee, and Karahallos]{eslami_user_2019}
M.~Eslami, A.~Elazari Bar~On, K.~Vaccaro, E.~Gilbert, M.~K. Lee, and K.~Karahallos.
\newblock User {Attitudes} towards {Algorithmic} {Opacity} and {Transparency} in {Online} {Reviewing} {Platforms}.
\newblock \emph{Proceedings of the 2019 CHI Conference on Human Factors in Computing Systems}, pages 1--14, 2019.

\bibitem[Fast and Horvitz(2017)]{fast_long-term_2017}
E.~Fast and E.~Horvitz.
\newblock Long-{Term} {Trends} in the {Public} {Perception} of {Artificial} {Intelligence}.
\newblock \emph{Proceedings of the AAAI Conference on Artificial Intelligence}, 31\penalty0 (1), Feb. 2017.
\newblock ISSN 2374-3468, 2159-5399.
\newblock \doi{10.1609/aaai.v31i1.10635}.
\newblock URL \url{https://ojs.aaai.org/index.php/AAAI/article/view/10635}.

\bibitem[French and Hancock(2017)]{french_whats_2017}
M.~French and J.~Hancock.
\newblock What's the {Folk} {Theory}? {Reasoning} {About} {Cyber}-{Social} {Systems}, Feb. 2017.
\newblock URL \url{https://papers.ssrn.com/abstract=2910571}.

\bibitem[Gebru and Torres(2024)]{gebru_tescreal_2024}
T.~Gebru and E.~P. Torres.
\newblock The {TESCREAL} bundle: {Eugenics} and the promise of utopia through artificial general intelligence.
\newblock \emph{First Monday}, Apr. 2024.
\newblock ISSN 1396-0466.
\newblock \doi{10.5210/fm.v29i4.13636}.
\newblock URL \url{https://firstmonday.org/ojs/index.php/fm/article/view/13636}.

\bibitem[Goertzel(2007)]{goertzel_human-level_2007}
B.~Goertzel.
\newblock Human-level artificial general intelligence and the possibility of a technological singularity: {A} reaction to {Ray} {Kurzweil}'s {The} {Singularity} {Is} {Near}, and {McDermott}'s critique of {Kurzweil}.
\newblock \emph{Artificial Intelligence}, 171\penalty0 (18):\penalty0 1161--1173, Dec. 2007.
\newblock ISSN 0004-3702.
\newblock \doi{10.1016/j.artint.2007.10.011}.
\newblock URL \url{https://www.sciencedirect.com/science/article/pii/S0004370207001464}.

\bibitem[Goldin et~al.(2024)Goldin, Koutroumpis, Lafond, and Winkler]{goldin_why_2024}
I.~Goldin, P.~Koutroumpis, F.~Lafond, and J.~Winkler.
\newblock Why {Is} {Productivity} {Slowing} {Down}?
\newblock \emph{Journal of Economic Literature}, 62\penalty0 (1):\penalty0 196--268, Mar. 2024.
\newblock ISSN 0022-0515.
\newblock \doi{10.1257/jel.20221543}.
\newblock URL \url{https://www.aeaweb.org/articles?id=10.1257/jel.20221543}.

\bibitem[Gray(2011)]{gray_road_2011}
J.~Gray.
\newblock On the {Road} to {Immortality}.
\newblock \emph{The New York Review of Books}, Nov. 2011.
\newblock URL \url{https://www.nybooks.com/articles/2011/11/24/road-immortality/?pagination=false}.

\bibitem[Guterres(2023)]{guterres_secretary-generals_2023}
A.~Guterres.
\newblock Secretary-{General}'s opening remarks at press briefing on {Policy} {Brief} on {Information} {Integrity} on {Digital} {Platforms}, June 2023.
\newblock URL \url{https://www.un.org/sg/en/content/sg/speeches/2023-06-12/secretary-generals-opening-remarks-press-briefing-policy-brief-information-integrity-digital-platforms}.

\bibitem[Heaven(2023)]{heaven_geoffrey_2023}
W.~D. Heaven.
\newblock Geoffrey {Hinton} tells us why he's now scared of the tech he helped build.
\newblock \emph{MIT Technology Review}, May 2023.
\newblock URL \url{https://www.technologyreview.com/2023/05/02/1072528/geoffrey-hinton-google-why-scared-ai/}.

\bibitem[Ho and Lee(2015)]{ho_typology_2015}
J.~C. Ho and C.-S. Lee.
\newblock A typology of technological change: {Technological} paradigm theory with validation and generalization from case studies.
\newblock \emph{Technological Forecasting and Social Change}, 97:\penalty0 128--139, Aug. 2015.
\newblock ISSN 0040-1625.
\newblock \doi{10.1016/j.techfore.2014.05.015}.
\newblock URL \url{https://www.sciencedirect.com/science/article/pii/S0040162514001863}.

\bibitem[Imbrie et~al.(2021)Imbrie, Gelles, Dunham, and Aiken]{imbrie_contending_2021}
A.~Imbrie, R.~Gelles, J.~Dunham, and C.~Aiken.
\newblock Contending {Frames}: {Evaluating} {Rhetorical} {Dynamics} in {AI}.
\newblock Technical report, Center for Security and Emerging Technology, May 2021.
\newblock URL \url{https://cset.georgetown.edu/publication/contending-frames/}.

\bibitem[Khatchadourian(2015)]{khatchadourian_doomsday_2015}
R.~Khatchadourian.
\newblock The {Doomsday} {Invention}: {Will} artificial intelligence bring us utopia or destruction?".
\newblock \emph{The New Yorker}, Nov. 2015.
\newblock URL \url{https://www.newyorker.com/magazine/2015/11/23/doomsday-invention-artificial-intelligence-nick-bostrom}.

\bibitem[Korinek and Stiglitz(2017)]{korinek_artificial_2017}
A.~Korinek and J.~E. Stiglitz.
\newblock Artificial {Intelligence} and {Its} {Implications} for {Income} {Distribution} and {Unemployment}, Dec. 2017.
\newblock URL \url{https://www.nber.org/papers/w24174}.

\bibitem[Korinek and Suh(2024)]{korinek_scenarios_2024}
A.~Korinek and D.~Suh.
\newblock Scenarios for the {Transition} to {AGI}.
\newblock Technical Report w32255, National Bureau of Economic Research, Cambridge, MA, Mar. 2024.
\newblock URL \url{http://www.nber.org/papers/w32255.pdf}.

\bibitem[Korotayev and LePoire(2020)]{korotayev_21st_2020}
A.~V. Korotayev and D.~J. LePoire, editors.
\newblock \emph{The 21st {Century} {Singularity} and {Global} {Futures}: {A} {Big} {History} {Perspective}}.
\newblock Springer, Cham, Switzerland, 1st ed. 2020 edition edition, Jan. 2020.
\newblock ISBN 978-3-030-33729-2.

\bibitem[Kurzweil(1990)]{kurzweil_age_1990}
R.~Kurzweil.
\newblock \emph{The {Age} of {Intelligent} {Machines}}.
\newblock MIT Press, Cambridge, Mass, 1990.
\newblock ISBN 0-262-11121-7.

\bibitem[Kurzweil(2001)]{kurzweil_singularity_2001}
R.~Kurzweil.
\newblock The {Singularity}: {A} {Talk} {With} {Ray} {Kurzweil}.
\newblock \emph{Edge}, Mar. 2001.
\newblock URL \url{https://www.edge.org/conversation/ray_kurzweil-the-singularity}.

\bibitem[Kurzweil(2005)]{kurzweil_singularity_2005}
R.~Kurzweil.
\newblock \emph{The {Singularity} {Is} {Near}: {When} {Humans} {Transcend} {Biology}}.
\newblock Penguin Books, New York, NY, Sept. 2005.
\newblock ISBN 978-0-14-303788-0.

\bibitem[Kurzweil(2024)]{kurzweil_singularity_2024}
R.~Kurzweil.
\newblock \emph{The {Singularity} is {Nearer}: {When} {We} {Merge} with {Al}}.
\newblock Viking, New York, 2024.
\newblock ISBN 978-0-399-56276-1.
\newblock OCLC: 1438926317.

\bibitem[LePoire(2021)]{lepoire_exploring_2021}
D.~LePoire.
\newblock Exploring {Time} {Patterns} in {Big} {History}.
\newblock In L.~E. Grinin, I.~V. Ilyin, and A.~V. Korotayev, editors, \emph{Globalistics and {Globalization} {Studies}: {Current} and {Future} {Trends} in the {Big} {History} {Perspective}}, pages 254--263. ‘Uchitel’ Publishing House, Volgograd, 2021.
\newblock URL \url{https://www.sociostudies.org/almanac/articles/exploring_time_patterns_in_big_history/}.

\bibitem[Long and Magerko(2020)]{long_what_2020}
D.~Long and B.~Magerko.
\newblock What is {AI} {Literacy}? {Competencies} and {Design} {Considerations}.
\newblock \emph{Proceedings of the 2020 Conference on Human Factors in Computing Systems}, pages 1--16, Apr. 2020.
\newblock \doi{https://doi.org/10.1145/3313831.3376727}.

\bibitem[McGinn(1982)]{mcginn_character_1982}
C.~McGinn.
\newblock \emph{The {Character} of {Mind}}.
\newblock Oxford University Press, Oxford [Oxfordshire ] ; New York, 1982.
\newblock ISBN 978-0-19-289159-4 978-0-19-219171-7.

\bibitem[McGinn(1989)]{mcginn_can_1989}
C.~McGinn.
\newblock Can {We} {Solve} the {Mind}--{Body} {Problem}?
\newblock \emph{Mind}, 98\penalty0 (391):\penalty0 349--366, 1989.
\newblock ISSN 0026-4423.
\newblock URL \url{https://www.jstor.org/stable/2254848}.
\newblock Publisher: [Oxford University Press, Mind Association].

\bibitem[Mitchell(2021)]{mitchell_why_2021}
M.~Mitchell.
\newblock Why {AI} is {Harder} {Than} {We} {Think}, Apr. 2021.
\newblock URL \url{http://arxiv.org/abs/2104.12871}.
\newblock arXiv:2104.12871 [cs].

\bibitem[Modis(2002)]{modis_forecasting_2002}
T.~Modis.
\newblock Forecasting the growth of complexity and change.
\newblock \emph{Technological Forecasting and Social Change}, 69\penalty0 (4):\penalty0 377--404, May 2002.
\newblock ISSN 00401625.
\newblock \doi{10.1016/S0040-1625(01)00172-X}.
\newblock URL \url{https://linkinghub.elsevier.com/retrieve/pii/S004016250100172X}.

\bibitem[Modis(2005)]{modis_end_2005}
T.~Modis.
\newblock The end of the internet rush.
\newblock \emph{Technological Forecasting and Social Change}, 72\penalty0 (8):\penalty0 938--943, Oct. 2005.
\newblock ISSN 0040-1625.
\newblock \doi{10.1016/j.techfore.2005.06.004}.
\newblock URL \url{https://www.sciencedirect.com/science/article/pii/S0040162505000843}.

\bibitem[Modis(2006)]{modis_singularity_2006}
T.~Modis.
\newblock The {Singularity} {Myth}, Dec. 2006.
\newblock URL \url{https://osf.io/3m9cn}.

\bibitem[Modis(2007)]{modis_strengths_2007}
T.~Modis.
\newblock Strengths and weaknesses of {S}-curves.
\newblock \emph{Technological Forecasting and Social Change}, 74:\penalty0 866--872, July 2007.
\newblock \doi{10.1016/j.techfore.2007.04.005}.

\bibitem[Modis(2012)]{modis_why_2012}
T.~Modis.
\newblock Why the {Singularity} {Cannot} {Happen}.
\newblock In A.~H. Eden, J.~H. Moor, J.~H. Søraker, and E.~Steinhart, editors, \emph{Singularity {Hypotheses}: {A} {Scientific} and {Philosophical} {Assessment}}, pages 311--346. Springer, Berlin, Heidelberg, 2012.
\newblock ISBN 978-3-642-32560-1.
\newblock \doi{10.1007/978-3-642-32560-1_16}.
\newblock URL \url{https://doi.org/10.1007/978-3-642-32560-1_16}.

\bibitem[Morris et~al.(2024)Morris, Sohl-Dickstein, Fledel, Warkentin, Dafoe, Faust, Farabet, and Legg]{morris_position_2024}
M.~R. Morris, J.~Sohl-Dickstein, N.~Fledel, T.~Warkentin, A.~Dafoe, A.~Faust, C.~Farabet, and S.~Legg.
\newblock Position: {Levels} of {AGI} for {Operationalizing} {Progress} on the {Path} to {AGI}.
\newblock \emph{Proceedings of the 41st International Conference on Machine Learning}, 2024.

\bibitem[Müller and Bostrom(2016)]{muller_future_2016}
V.~C. Müller and N.~Bostrom.
\newblock Future {Progress} in {Artificial} {Intelligence}: {A} {Survey} of {Expert} {Opinion}.
\newblock In V.~C. Müller, editor, \emph{Fundamental {Issues} of {Artificial} {Intelligence}}, pages 555--572. Springer International Publishing, Cham, 2016.
\newblock ISBN 978-3-319-26485-1.
\newblock \doi{10.1007/978-3-319-26485-1_33}.
\newblock URL \url{https://doi.org/10.1007/978-3-319-26485-1_33}.

\bibitem[Noy and Zhang(2023)]{noy_experimental_2023}
S.~Noy and W.~Zhang.
\newblock Experimental evidence on the productivity effects of generative artificial intelligence.
\newblock \emph{Science}, 381\penalty0 (6654):\penalty0 187--192, July 2023.
\newblock ISSN 0036-8075, 1095-9203.
\newblock \doi{10.1126/science.adh2586}.
\newblock URL \url{https://www.science.org/doi/10.1126/science.adh2586}.

\bibitem[O'Gieblyn(2021)]{ogieblyn_god_2021}
M.~O'Gieblyn.
\newblock \emph{God, {Human}, {Animal}, {Machine}: {Technology}, {Metaphor}, and the {Search} for {Meaning}}.
\newblock Vintage, New York, Aug. 2021.
\newblock ISBN 978-0-525-56271-9.

\bibitem[Peng et~al.(2023)Peng, Kalliamvakou, Cihon, and Demirer]{peng_impact_2023}
S.~Peng, E.~Kalliamvakou, P.~Cihon, and M.~Demirer.
\newblock The {Impact} of {AI} on {Developer} {Productivity}: {Evidence} from {GitHub} {Copilot}, Feb. 2023.
\newblock URL \url{http://arxiv.org/abs/2302.06590}.
\newblock arXiv:2302.06590 [cs].

\bibitem[Phillips(2007)]{phillips_s-curves_2007}
F.~Phillips.
\newblock On {S}-curves and tipping points.
\newblock \emph{Technological Forecasting and Social Change}, 74\penalty0 (6):\penalty0 715--730, July 2007.
\newblock ISSN 0040-1625.
\newblock \doi{10.1016/j.techfore.2006.11.006}.
\newblock URL \url{https://www.sciencedirect.com/science/article/pii/S0040162506002125}.

\bibitem[Phillips and Linstone(2016)]{phillips_key_2016}
F.~Phillips and H.~Linstone.
\newblock Key ideas from a 25-year collaboration at technological forecasting \& social change.
\newblock \emph{Technological Forecasting and Social Change}, 105:\penalty0 158--166, Apr. 2016.
\newblock ISSN 0040-1625.
\newblock \doi{10.1016/j.techfore.2016.01.007}.
\newblock URL \url{https://www.sciencedirect.com/science/article/pii/S0040162516000081}.

\bibitem[Pohl and Goldkind(2023)]{pohl_ai_2023}
B.~Pohl and L.~Goldkind.
\newblock {AI} folk tales: how nontechnical publics make sense of artificial intelligence.
\newblock In S.~Nah, editor, \emph{Research handbook on artificial intelligence and communication}, pages 246--266. Edward Elgar Publishing, Cheltenham, UK, 2023.
\newblock URL \url{https://acrobat.adobe.com/id/urn:aaid:sc:VA6C2:a4ccddec-1672-428d-af61-dd74a46adbce}.

\bibitem[Pouget et~al.(2024)Pouget, Dennis, Bateman, Trager, Araujo, Belfield, Cleeland, Estier, Futerman, Guest, Ignacio, Kannan, Maas, Mahoney, Martinet, Mökander, Ng, Ó~hÉigeartaigh, Peppin, Seifert, Singer, Stauffer, Whithers, and Ziosi]{pouget_future_2024}
H.~Pouget, C.~Dennis, J.~Bateman, R.~Trager, R.~Araujo, H.~Belfield, B.~Cleeland, M.~Estier, G.~Futerman, O.~Guest, C.~Ignacio, V.~Kannan, M.~Maas, C.~Mahoney, C.~Martinet, J.~Mökander, K.~Y. Ng, S.~Ó~hÉigeartaigh, A.~Peppin, K.~Seifert, S.~Singer, M.~Stauffer, C.~Whithers, and M.~Ziosi.
\newblock The {Future} of {International} {Scientific} {Assessments} of {AI}’s {Risks}.
\newblock Technical report, Carnegia Endowment for International Peace, Washington, DC, Aug. 2024.

\bibitem[Ptolemy(2011)]{ptolemy_transcendent_2011}
B.~Ptolemy.
\newblock Transcendent {Man}, 2011.

\bibitem[Rothman(2023)]{rothman_why_2023}
J.~Rothman.
\newblock Why the {Godfather} of {A}.{I}. {Fears} {What} {He}'s {Built}.
\newblock \emph{The New Yorker}, Nov. 2023.
\newblock URL \url{https://www.newyorker.com/magazine/2023/11/20/geoffrey-hinton-profile-ai}.

\bibitem[Sandberg(2013)]{more_overview_2013}
A.~Sandberg.
\newblock An {Overview} of {Models} of {Technological} {Singularity}.
\newblock In M.~More and N.~Vita‐More, editors, \emph{The {Transhumanist} {Reader}}, pages 376--394. Wiley, 1 edition, Apr. 2013.
\newblock ISBN 978-1-118-33429-4 978-1-118-55592-7.
\newblock \doi{10.1002/9781118555927.ch36}.
\newblock URL \url{https://onlinelibrary.wiley.com/doi/10.1002/9781118555927.ch36}.

\bibitem[Sartori and Bocca(2023)]{sartori_minding_2023}
L.~Sartori and G.~Bocca.
\newblock Minding the gap(s): public perceptions of {AI} and socio-technical imaginaries.
\newblock \emph{AI \& Society}, 38\penalty0 (2):\penalty0 443--458, Apr. 2023.
\newblock ISSN 0951-5666, 1435-5655.
\newblock \doi{10.1007/s00146-022-01422-1}.
\newblock URL \url{https://link.springer.com/10.1007/s00146-022-01422-1}.

\bibitem[Searle(1980)]{searle_minds_1980}
J.~R. Searle.
\newblock Minds, brains, and programs.
\newblock \emph{Behavioral and Brain Sciences}, 3\penalty0 (3):\penalty0 417--424, Sept. 1980.
\newblock ISSN 1469-1825, 0140-525X.
\newblock \doi{10.1017/S0140525X00005756}.
\newblock URL \url{https://www.cambridge.org/core/journals/behavioral-and-brain-sciences/article/minds-brains-and-programs/DC644B47A4299C637C89772FACC2706A}.

\bibitem[Searle(1983)]{searle_intentionality_1983}
J.~R. Searle.
\newblock \emph{Intentionality, an essay in the philosophy of mind}.
\newblock Cambridge paperback library. Cambridge University Press, Cambridge [Cambridgeshire], 1983.
\newblock ISBN 978-0-521-22895-4.
\newblock URL \url{http://catdir.loc.gov/catdir/toc/cam029/82019849.html}.
\newblock OCLC: 9196773.

\bibitem[Searle(1986)]{searle_minds_1986}
J.~R. Searle.
\newblock \emph{Minds, {Brains} and {Science}}.
\newblock Harvard University Press, Jan. 1986.
\newblock ISBN 978-0-674-26721-3.
\newblock Google-Books-ID: Q8fmEAAAQBAJ.

\bibitem[Shanahan(2004)]{shanahan_frame_2004}
M.~Shanahan.
\newblock The {Frame} {Problem}, Feb. 2004.
\newblock URL \url{https://plato.stanford.edu/ENTRIES/frame-problem/}.
\newblock Last Modified: 2016-02-08.

\bibitem[Shumailov et~al.(2024)Shumailov, Shumaylov, Zhao, Gal, Papernot, and Anderson]{shumailov_curse_2024}
I.~Shumailov, Z.~Shumaylov, Y.~Zhao, Y.~Gal, N.~Papernot, and R.~Anderson.
\newblock The {Curse} of {Recursion}: {Training} on {Generated} {Data} {Makes} {Models} {Forget}, 2024.
\newblock URL \url{https://arxiv.org/abs/2305.17493}.
\newblock Version Number: 3.

\bibitem[Singler(2018)]{singler_rokos_2018}
B.~Singler.
\newblock Roko’s {Basilisk} or {Pascal}’s? {Thinking} of {Singularity} {Thought} {Experiments} as {Implicit} {Religion}.
\newblock \emph{Implicit Religion}, 20\penalty0 (3):\penalty0 279--297, May 2018.
\newblock ISSN 1743-1697, 1463-9955.
\newblock \doi{10.1558/imre.35900}.
\newblock URL \url{https://journal.equinoxpub.com/IR/article/view/3226}.

\bibitem[Snooks(2020)]{snooks_is_2020}
G.~D. Snooks.
\newblock Is {Singularity} a {Scientific} {Concept} or the {Construct} of {Metaphysical} {Historicism}? {Implications} for {Big} {History}.
\newblock In A.~V. Korotayev and D.~J. LePoire, editors, \emph{The 21st {Century} {Singularity} and {Global} {Futures}: {A} {Big} {History} {Perspective}}, pages 225--263. Springer International Publishing, Cham, 2020.
\newblock ISBN 978-3-030-33730-8.
\newblock \doi{10.1007/978-3-030-33730-8_12}.
\newblock URL \url{https://doi.org/10.1007/978-3-030-33730-8_12}.

\bibitem[Suleyman(2023)]{suleyman_coming_2023}
M.~Suleyman.
\newblock \emph{The {Coming} {Wave}: {Technology}, {Power}, and the {Twenty}-first {Century}'s {Greatest} {Dilemma}}.
\newblock Crown, New York, Sept. 2023.
\newblock ISBN 978-0-593-59395-0.

\bibitem[Toner(2024)]{toner_oversight_2024}
H.~Toner.
\newblock Oversight of {AI}: {Insiders}’ {Perspectives}, {Testimony} {Before} the {U}.{S}. {Senate} {Committee} on the {Judiciary}, Sept. 2024.
\newblock URL \url{https://www.judiciary.senate.gov/committee-activity/hearings/oversight-of-ai-insiders-perspectives}.

\bibitem[Trammell and Korinek(2023)]{trammell_economic_2023}
P.~Trammell and A.~Korinek.
\newblock Economic {Growth} under {Transformative} {AI}, Oct. 2023.
\newblock URL \url{https://www.nber.org/papers/w31815}.

\bibitem[Ulam(1958)]{ulam_tribute_1958}
S.~Ulam.
\newblock Tribute to {John} von {Neumann}.
\newblock \emph{Bulletin of the American Mathematical Society}, 64\penalty0 (3):\penalty0 1--49, 1958.

\bibitem[Vinge(1983)]{vinge_first_1983}
V.~Vinge.
\newblock First {Word}.
\newblock \emph{Omni}, page~10, Jan. 1983.

\bibitem[Vinge(1993)]{vinge_coming_1993}
V.~Vinge.
\newblock The coming technological singularity: {How} to survive in the post-human era.
\newblock In \emph{VISION-21 Symposium}, Dec. 1993.
\newblock URL \url{https://ntrs.nasa.gov/citations/19940022856}.
\newblock NTRS Author Affiliations: San Diego State Univ. NTRS Document ID: 19940022856 NTRS Research Center: Legacy CDMS (CDMS).

\bibitem[Ytre-Arne and Moe(2021)]{ytre-arne_folk_2021}
B.~Ytre-Arne and H.~Moe.
\newblock Folk theories of algorithms: {Understanding} digital irritation.
\newblock \emph{Media, Culture \& Society}, 43\penalty0 (5):\penalty0 807--824, July 2021.
\newblock ISSN 0163-4437, 1460-3675.
\newblock \doi{10.1177/0163443720972314}.
\newblock URL \url{http://journals.sagepub.com/doi/10.1177/0163443720972314}.

\bibitem[Yudkowsky(2007)]{yudkowsky_three_2007}
E.~Yudkowsky.
\newblock Three {Major} {Singularity} {Schools}, Sept. 2007.
\newblock URL \url{https://intelligence.org/2007/09/30/three-major-singularity-schools/}.

\end{thebibliography}

\end{document}